\documentclass[%
 aip,
 amsmath,amssymb,
 reprint,%
]{revtex4-1}

\usepackage{graphicx}
\usepackage{dcolumn}
\usepackage{bm}

\usepackage[utf8]{inputenc}
\usepackage[T1]{fontenc}
\usepackage{mathptmx}

\usepackage{graphicx}
\usepackage{dcolumn}
\usepackage{bm}
\usepackage[utf8]{inputenc}
\usepackage[T1]{fontenc}
\usepackage{amsmath,amssymb,amsfonts,graphics,graphicx,dcolumn,bm,enumerate,amsthm}
\usepackage{comment,natbib,appendix,mathtools}

\begin{document}

\preprint{AIP/123-QED}

\title{Phase transition in the Kolkata Paise Restaurant problem}

\author{Antika Sinha}
\email{antikasinha@gmail.com}
 \affiliation{Department of Computer Science, Asutosh College, Kolkata-700026, India}

 \author{Bikas K. Chakrabarti}
 \email{bikask.chakrabarti@saha.ac.in}
\affiliation{Saha Institute of Nuclear Physics, Kolkata-700064, India}
\affiliation{S.N.Bose National Centre for Basic Sciences, Kolkata-700106, India}
\affiliation{Economic Research Unit, Indian Statistical Institute, Kolkata-700108, India}

\date{\today}

\begin{abstract}
A novel phase transition behaviour is observed in the Kolkata Paise Restaurant (KPR) problem where large number ($N$) of agents or customers collectively (and iteratively) learn to choose among the $N$ restaurants where she would expect to be alone that evening and would get the only dish available there (or may get randomly picked up if more than one agent arrive there that evening). The players are expected to evolve their strategy such that the publicly available information about past crowd in different restaurants can be utilized and each of them is able to make the best minority choice. For equally ranked restaurants we follow two crowd-avoiding strategies: Strategy I, where each of the $n_i(t)$ number of agents arriving at the $i$-th restaurant on the $t$-th evening goes back to the same restaurant on the next evening with probability $[n_i(t)]^{-\alpha}$, while in Strategy II, with probability $p$, when $n_i(t) > 1$. We study the steady state ($t$-independent) utilization fraction $f:(1-f)$ giving the steady state (wastage) fraction of restaurants going without any customer in any particular evening. With both the strategies we find, near $\alpha_c=0_+$ (in strategy I) or $p=1_-$ (in strategy II), the steady state wastage fraction $(1-f)\propto(\alpha - \alpha_c)^{\beta}$ or $(p_c - p)^\beta$ with $\beta \simeq 0.8, 0.87, 1.0$ and the convergence time $\tau$ (for $f(t)$ becoming independent of $t$) varies as $\tau\propto{(\alpha-\alpha_c)}^{-\gamma}$ or ${(p_c-p)}^{-\gamma}$, with $\gamma \simeq 1.18, 1.11, 1.05$ in infinite-dimension (rest of the $N-1$ neighboring restaurants), three-dimension ($6$ neighbors) and two-dimension ($4$ neighbors) respectively.

\end{abstract}

\maketitle

\begin{quotation}
Social games where the players or agents try to choose for the less crowded or minority solutions (to avail of the scarce resources) are very common. In such games, a macroscopically large number of agents make decision parallelly and iteratively (in absence of any dictator), based on publicly available information (regarding past mistakes and successes), to choose where she can be alone (and avail the resource). Eventually, such collective learning makes it socially efficient. One such social minority game is the Kolkata Paise Restaurant (KPR) problem. We study the steady state statistics and the phase transition behaviour of the KPR problem \cite{chakrabarti2009kolkata,chakrabarti2017econophysics}. KPR is a many-agent and many-choice repeated game, where the agents collectively learn from past mistakes, how to share best the limited resources. In this kind of games, each agent tries to anticipate and choose her own strategy, every time (learning from the publicly available past informations) in parallel mode (unguided; in absence of any instruction or non-playing agent/dictator). The restaurants are assumed to prepare every evening fixed meal plates which are equally priced (hence no budget constraint for customers). Only the crowd avoidance abilities determines the individual success in securing meal on any or successive evenings.
We show that in KPR, two stochastic strategies can eventually lead to the most efficient social solution at some limiting control parameter values, corresponding to a phase transition point, thereby implying very long convergence time (critical slowing down). 

\end{quotation}

\section{Introduction}

Specifically, we consider here the case of $N$ restaurants and $N$ agents or customers or players who decide every evening (on the basis of informations about the past evenings, available to everyone), which restaurant to choose such that she will be alone there and will get the meal. Each restaurant is assumed, for simplicity, to prepare one dish every evening (generalization does not help getting any further insight at this stage). For more than one person arriving any restaurant any evening, a randomly chosen one will get the meal and rest (arriving there) will not get any that evening.

Although every evening each of the $N$ restaurants prepares one dish and in-principle everyone is entitled to a dish every evening, overcrowding due to stochasticity of choices make the probability of success for each customer less than unity in such (democratic choice) games. We measure the social efficiency by the utilization fraction $f(t)$ on any day (evening) $t$ as
\begin{equation}\label{eq1}
f(t) = [1-\sum_{i=1}^{N} [\delta(n_i(t))/N]]
\end{equation}

\noindent
with $\delta(n) = 1,0$ for $n = 0,\geq 1$ respectively; $n_i(t)$ denotes the number of agents arriving at the $i$-th (rank) restaurant on $t$-th evening. The fraction $(1-f(t))$ gives the fraction of social wastage or the fraction of restaurants going without any customer on the $t$-th evening. The objective of social learning strategies here is to achieve $f(t)$ = 1 preferably in finite convergence time ($\tau$), i.e., for $t$ $\geq \tau$, or at least as $t\to\infty$ (see e.g., \cite{chakrabarti2017econophysics,ghosh2010statistics,ghosh2010kolkata}).

Indeed, a dictated solution is extremely simple and very efficient: the dictator asks everyone to form a queue and visit the restaurants according to their respective positions in the queue and then asks them to shift their positions by one step (rank) in the next evening (assuming periodic boundary condition). Everyone gets the food: No wastage, i.e., the steady state ($t$-independent) utilization fraction $f=1$, and that too from the first evening (convergence time $\tau$ is zero). This is true even when the restaurants have ranks (agreed by all the agents or customers). However, in reality (in democracy), this dictated solution is not acceptable and each agent would like to (learn from past experience and) decide on her own every evening which restaurant to choose such that she is alone there and gets the dish. The more successful such collective learning, the more is the utilization fraction of the services. Question is, what is the maximum utilization fraction value ($f$) and convergence time ($\tau$) of such  `learned' democratic choices (due to individually learned and chosen strategies) for a large society ($N\to\infty$). Note that the dictated solution gives full utilization ($f$ = 1) and that too in zero convergence time ($\tau$ = 0) for any $N$.

Assuming that no past history of restaurant occupancy is available, i.e., no learning, let us consider the process of randomly choosing any  of the $N$ restaurant by $\lambda N$ agents (we consider $\lambda$ = 1 in KPR game later). Then the probability of choosing any restaurant by $m ~(>1)$ agents on any evening is 
\begin{subequations}\label{eq2}
\begin{gather}
  \rho(m) = \binom{\lambda N}{m} {p}^{m} {(1-p)}^{\lambda N - m} \label{eq2.1} \\
\shortintertext{where $p = \frac{1}{N}$; giving $\rho(m)$ = $[\lambda^m /(m!)]~exp(-\lambda)$ as $N \to \infty$. Hence \cite{chakrabarti2009kolkata} the average fraction of restaurants not chosen on any evening is $\rho(m=0)=exp(-\lambda)$, and average fraction of restaurants filled or utilized on any evening (Eq.~\ref{eq1}), is given by} 
f=1-exp(-\lambda)\simeq 0.63,\:\textrm{for}\:\lambda = 1. \label{eq2.2}
\end{gather}
\end{subequations}

As we mentioned earlier, the agents would try to learn from the past mistakes in making their respective choices and can improve her chance to be in the minority with efficient learning strategy. We study here the dynamics of the game with two such stochastic learning strategies which allow for considerable increase in the eventual (steady state) value of the utilization fraction ($f$). Specifically, we study here two strategies (Strategy I and Strategy II), giving the probabilities to choose going back or not to the same restaurant visited last evening or to another restaurant, depending on the last evening's crowd-size in the chosen restaurant. We find interesting phase transition behaviour (with identical universality class) with both of these strategies. This transition behaviour is qualitatively different from the transitions observed \cite{ghosh2012phase,ghosh2014zipf} earlier with agents' sticking probability to any chosen restaurant and with limited resources ($\lambda < 1$) in KPR and similar models.

We studied here the phase transition behaviour (from steady state value of $f<1$ to $f=1$) and convergence or relaxation time $\tau$, diverging as the critical point is approached, using both the strategies I and II in different dimensions where the choice of the agent to shift to the neighboring restaurants next evening are different (to any of the other $(N-1)$ restaurants in $d\to\infty$, to neighboring six restaurants in $d=3$ and to four neighboring restaurants in $d=2$).

\section{Strategy Description}
In the following, we study numerically the dynamics of KPR game played by $N$ agents (interchangeably called players or agents) such that each evening (interchangeably called day or time) each of the agents employ some stochastic strategy, based on the past crowd information in different restaurants (available to each players), helping her to choose among $N$ restaurants, maximizing her chance of arriving at a vacant restaurant that evening and to get the dish. We study here the following strategies: I and II. We consider here all the restaurants as equally ranked (none is preferred more than the other and the choice depends only on the past crowd size).

\subparagraph{Strategy I:}
The strategy here is that any agent tries to go back to the same restaurant as chosen in the earlier evening (day) with a probability decreasing with an inverse power of the crowd size arriving there last evening and goes to any other restaurant randomly with the rest of the probability. In other words, on day $t$, an agent goes back to her last day's visited restaurant $k$ with probability 
\begin{subequations}
\begin{gather}
  p^{(I)}_k(t) = {[n_k(t-1)] }^{-\alpha},\:\alpha > 0. \label{eq3.1}
\shortintertext{If she was one of the $n_k(t-1)$ agents or players arriving there ($k$-th restaurant) last day $t-1$, she chooses one $(k^{'}\neq k)$  among any of the neighboring restaurants $n_r$ ($n_r = (N-1)$ in $\infty d$, 6 in $3d$, 4 in $2d$, 2 in $1d$ lattice structure) on day $t$, with probability }
p^{(I)}_{k^{'}}(t) = (1-p^{(I)}_k(t))/n_r. \label{eq3.2}
\end{gather}
\end{subequations}

\begin{figure*}
\resizebox{2.0\columnwidth}{!}{
\includegraphics{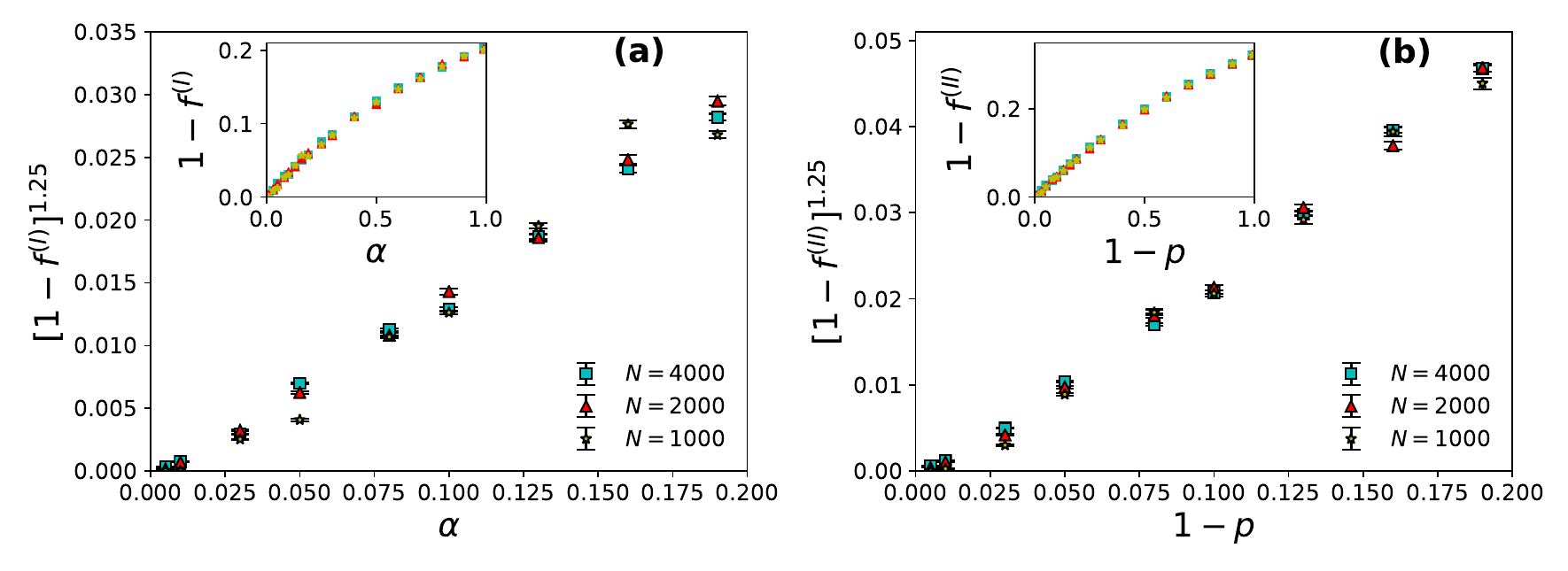}
}

\caption{Results for infinite dimension. (a) Plot of $(1-f^{(I)})$ against $\alpha$ and (b) plot of $(1-f^{(II)})$ against $1-p$ ($f$ denoting steady state social utilization fraction), show that $(1-f)$ variations fit well with power laws for $\alpha$ or $(1-p)$. Specifically we found that for $\alpha$ or $p$ values near $\alpha_c(=0_+)$ for strategy I and near $p_c(=1_-)$ for strategy II, power law holds for $(1-f^{(I)})\sim(\alpha - \alpha_c)^{\beta}\sim(1-f^{(II)})$ with $\beta = 0.80\pm0.05$. The insets show direct relationship between variations of $(1-f^{(I)})$ against $\alpha$ (in strategy I) and $(1-f^{(II)})$ with $p$ (in strategy II).
}\label{fig_inf_f}
\end{figure*}

\begin{figure*}
\resizebox{2.0\columnwidth}{!}{
\includegraphics{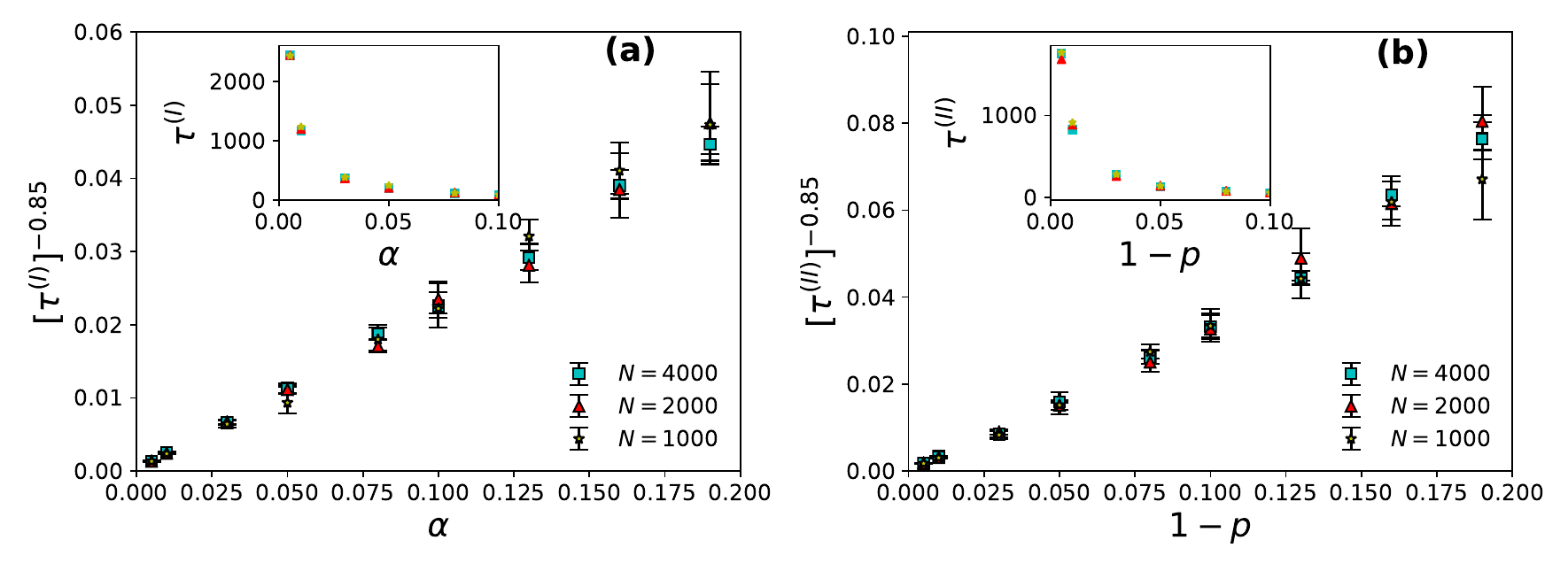}
}
\caption{Results for infinite dimension. (a) Plot of $\tau^{(I)}$ against $\alpha$ and (b) plot of $\tau^{(II)}$ against $1-p$ ($\tau$ denoting convergence or relaxation time taken to attain steady state social utilization fraction following learning Strategies I and II), showing that $\tau$ varies as power of $\alpha$ or $(1-p)$. Specifically, we found that for $\alpha$ or $p$ values near  $\alpha_c(=0_+)$ for strategy I and near $p_c(=1_-)$ for strategy II, a power law holds for $\tau^{(I)}\sim(\alpha - \alpha_c)^{-\gamma}\sim\tau^{(II)}\sim(p - p_c)^{-\gamma}$ where $\gamma = 1.18\pm0.05$. The insets show direct relationship between $\tau^{(I)}$ and $\alpha$ (in strategy I) and $\tau^{(II)}$ and $p$ (in strategy II).}
\label{fig_inf_t}
\end{figure*}

\subparagraph{Strategy II:}
We also consider the strategy where, on the $t$-th evening, every agent tries to go back to the same restaurant as chosen on the earlier evening with probability
\begin{subequations}
\begin{gather}
p^{(II)}_k(t) = 1,\:\textrm{if}\;n_k(t-1) = 1 \label{eq4.1}
\;\:\textrm{and}\\
p^{(II)}_{k^{'}}(t) = p < 1,\:\textrm{if}\;n_k(t-1) > 1 \label{eq4.2}
\end{gather}
\end{subequations}
for choosing any of the $n_r$ neighbouring restaurants restaurant ($k^{'} \neq k$).

Note that for $\alpha=0$ (in Strategy I) or for $p=1$ (in Strategy II), every evening she will return to the first day's chosen and visited restaurant, when of course the dynamics on successive days become trivial for both the strategies I and II. However $\alpha$ = $0_+$ (for Strategy I) or $p=1_-$ (for Strategy II) case can be extremely non-trivial and, as we will see, have interesting transition and other behaviours.

\section{Numerical Results}
Here we numerically study, with minimum $N$ = $1000$ (and maximum $N = 40^3$) and $\lambda = 1$ (number of restaurants = $N$ = number of agents/players), processing choice responses from each of $N$ agents to measure social utilization fraction $f$. Steady state occurs when $f$ becomes time independent. The steady state convergence time $\tau$ (for $t \geq \tau$, is when social utilization fraction $f$ does not change on average over the next hundred iteration, within a predefined error margin) is measured in units of time/iteration where each iteration corresponds to scanning as when each of the $N$ agents finishes one exercise of choices following Strategy I or II. Depending on the values of $\alpha$ or $p$ in strategy I (Eqs.~\ref{eq3.1},~\ref{eq3.2}) or II (Eqs.~\ref{eq4.1},~\ref{eq4.2}) respectively, values of the aggregated utilization fraction $f^{(I)}$ and $f^{(II)}$ (estimated using Eq.~\ref{eq1}) becomes unity for $\alpha\to\alpha_c$ (in strategy I) or for $p\to{p_c}$ (for strategy II), where $\alpha_c=0_+$ or $p_c \to 1_-$. We also study the growth of convergence time $\tau$ as $\alpha\to\alpha_c$ (for strategy I) or $p\to{p_c}$ (for strategy II) considering distinct lattices sizes. The restaurants are assumed to be situated on the lattice sites. Neighbour at each lattice site directs in $3d$, $2d$, $1d$ on both side(s) of each lattice direction of the present occupied restaurant. However in $\infty d$, every other restaurant is a nearest neighbor. Here, in order to avoid crowd of the last evening $(t-1)$, the agents can choose one among $n_r$ neighbouring restaurants following Eqs.~\ref{eq3.1},~\ref{eq3.2} for Strategy I or Eqs.~\ref{eq4.1},~\ref{eq4.2} for Strategy II on the next evening $t$. Note that since plotting a diverging function (here the convergence time $\tau$) is difficult, we plot instead it's inverse $(1/\tau)$ which tends to vanish at the same divergence point with an identical but negative exponent value.

\begin{figure*}
\resizebox{2.0\columnwidth}{!}{
\includegraphics{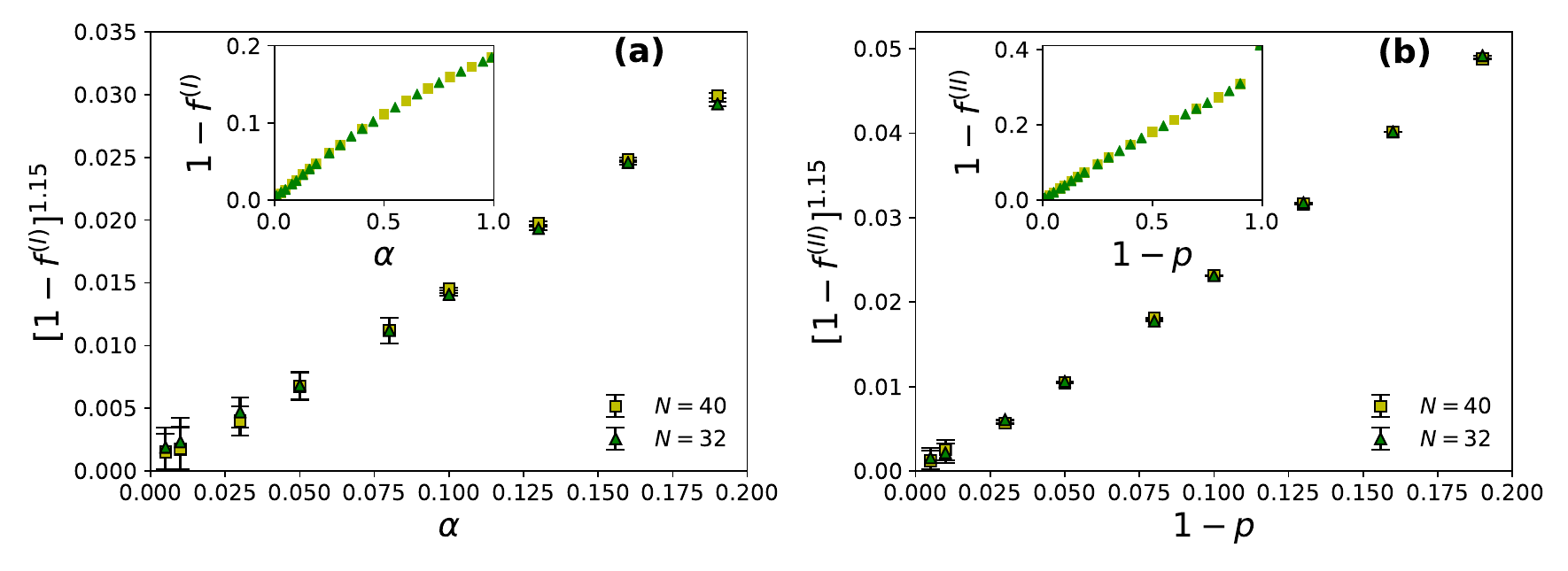}
}
\caption{Results for three dimension. (a) Plot of $(1-f^{(I)})$ against $\alpha$ and (b) plot of $(1-f^{(II)})$ against $1-p$ ($f$ being the steady state social utilization fraction), show that fitting of $(1-f)$ variations follow power laws for $\alpha$ or $(1-p)$. Specifically we found that for $\alpha$ or $p$ values near  $\alpha_c(=0_+)$ for strategy I and near $p_c(=1_-)$ for strategy II, power law holds for $(1-f^{(I)})\sim(\alpha - \alpha_c)^{\beta}\sim(1-f^{(II)})$ with $\beta = 0.87\pm0.05$. The insets show direct relationship between $(1-f^{(I)})$ and $\alpha$ (in strategy I) and $(1-f^{(II)})$ and $p$ (in strategy II) considering restaurants to be arranged on a simple cubic lattice (each restaurant having six neighbouring restaurants).}
\label{fig_3d_f}
\end{figure*}

\begin{figure*}
\resizebox{2.0\columnwidth}{!}{
\includegraphics{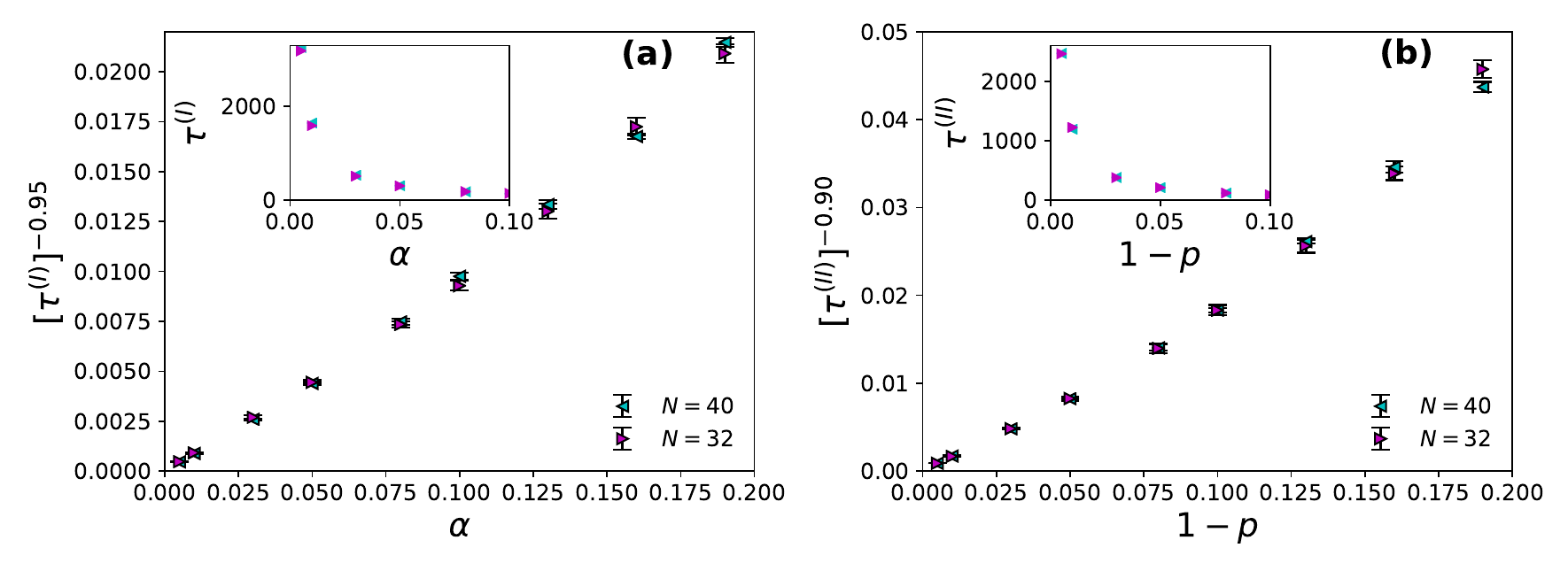}
}
\caption{Results for three dimension. (a) Plot of $\tau^{(I)}$ against $\alpha$ and (b) plot of $\tau^{(II)}$ against $1-p$ ($\tau$ denoting convergence or relaxation time taken to attain steady state social utilization fraction following learning Strategies I and II), representing that variations of $\tau$ fits well with power of $\alpha$ or $(1-p)$. Specifically, we found that for $\alpha$ or $p$ values near  $\alpha_c(=0_+)$ for strategy I and near $p_c(=1_-)$ for strategy II, a power law holds for $\tau^{(I)}\sim(\alpha - \alpha_c)^{-\gamma}\sim\tau^{(II)}\sim(p - p_c)^{-\gamma}$ where $\gamma = 1.11\pm0.05$. The insets show direct relationship between $\tau^{(I)}$ and $\alpha$ (in strategy I) and $\tau^{(II)}$ and $p$ (in strategy II).}
\label{fig_3d_t}
\end{figure*}

\subsection{Infinite dimensional lattice}
Here, if $n_i(t-1)$ number of agents had chosen the $i$-th restaurant last evening (time $t-1$) then each of them chooses to go back to the same $i$-th restaurant with probability $[n_i(t-1)]^{-\alpha}$ on the next ($t$-th) evening and chooses any of the remaining $N-1$ restaurants otherwise, for strategy I. With strategy II, the probability for each of the $n_i(t-1)$ agents to go back to the $i$-th restaurant on the next ($t$-th) evening is $p$ for $n_i(t-1)>1$ and to any other of the $N-1$ restaurants with probability $1-p$. After the system stabilizes i.e., when the utilization fraction $f(t)$ becomes independent of $t$, we note the steady state utilization fraction $f^{(I)}$ with strategy I or $f^{(II)}$ with strategy II and note the convergence time $\tau^{(I)}$ for strategy I and $\tau^{(II)}$ for strategy II, when $f(t)$ becomes $t$-independent.

We find the power law fits for wastage fraction $(1-f^{(I)})\sim(1-f^{(II)})\sim(\alpha - \alpha_c)^{\beta}\sim(p_c - p)^{\beta}$ with $\beta = 0.80 \pm 0.05$ (see Fig.~\ref{fig_inf_f}) and $\tau^{(I)}\sim\tau^{(II)}\sim(\alpha-\alpha_c)^{-\gamma}\sim (p_c - p)^{-\gamma}$ with $\gamma =  1.18 \pm 0.07$ (see Fig.~\ref{fig_inf_t}) for both of the strategies I and II. All simulations are done with $N$ up to $4000$ and the steady averages up to a maximum number of runs (days/evenings) of order $5\times10^5$. For estimating $\tau^{(I)}$ values, we looked for the convergence time $(\tau^{(I)})$ for $f^{(I)}(t)$ or $f^{(II)}(t)$ to attain the steady state value $f^{(I)}$ (within a small fore-assigned error margin).

\begin{figure*}
\resizebox{2.0\columnwidth}{!}{
\includegraphics{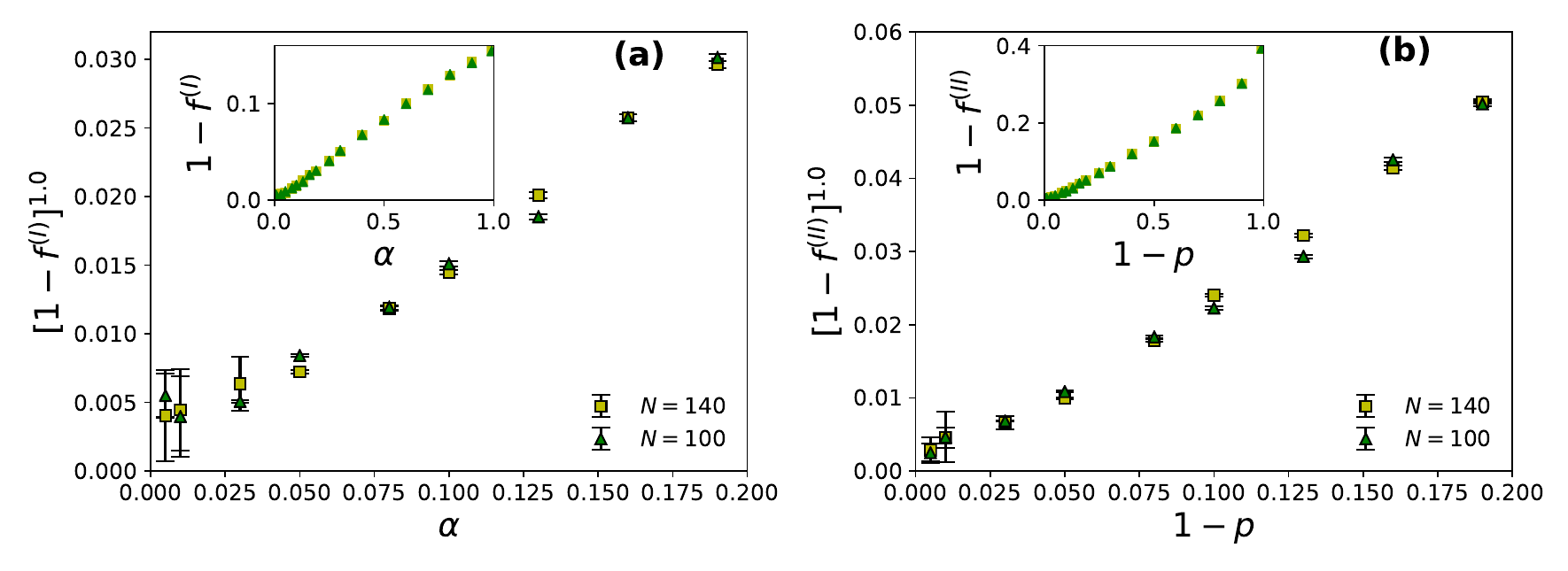}
}
\caption{Results for two dimension. (a) Plot of $(1-f^{(I)})$ against $\alpha$ and (b) plot of $(1-f^{(II)})$ against $1-p$ ($f$ representing the steady state social utilization fraction), show that $(1-f)$ variations fit as power law for $\alpha$ or $(1-p)$. Specifically we found that for $\alpha$ or $p$ values near  $\alpha_c(=0_+)$ for strategy I and near $p_c(=1_-)$ for strategy II, power law holds for $(1-f^{(I)})\sim(\alpha - \alpha_c)^{\beta}\sim(1-f^{(II)})$ with $\beta = 1.0\pm0.05$. The insets show direct relationship between $(1-f^{(I)})$ and $\alpha$ (in strategy I) and $(1-f^{(II)})$ and $p$ (in strategy II) considering restaurants to be arranged on a square lattice (each restaurant having four neighbouring restaurants). All these studies are for $N<140\times140$ and averaged over $16\times10^{5}$ runs.}

\label{fig_2d_f}
\end{figure*}

\begin{figure*}
\resizebox{2.0\columnwidth}{!}{
\includegraphics{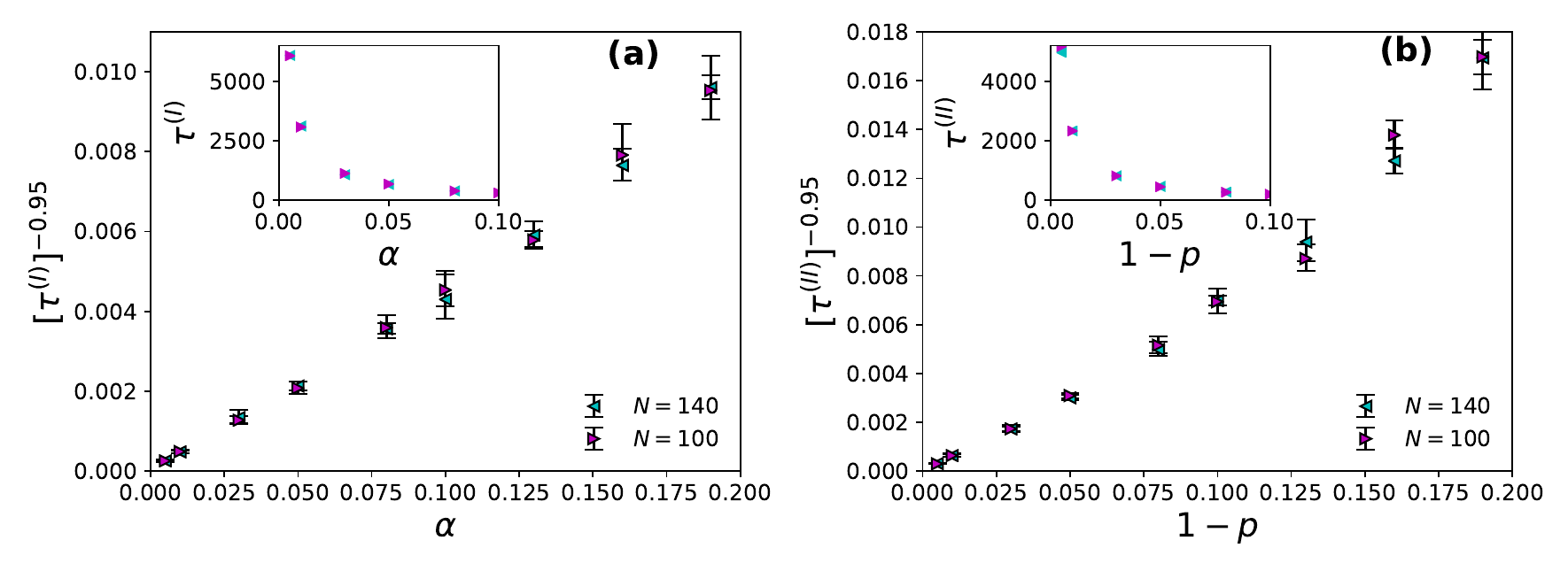}
}
\caption{Results for two dimension. (a) Plot of $\tau^{(I)}$ against $\alpha$ and (b) plot of $\tau^{(II)}$ against $1-p$ ($\tau$ being convergence or relaxation time taken to reach steady state social utilization fraction following learning Strategies I and II), representing that variations of $\tau$ fits well with power of $\alpha$ or $(1-p)$. Specifically, we found that for $\alpha$ or $p$ values near  $\alpha_c(=0_+)$ for strategy I and near $p_c(=1_-)$ for strategy II, a power law holds for $\tau^{(I)}\sim(\alpha - \alpha_c)^{-\gamma}\sim\tau^{(II)}\sim(p - p_c)^{-\gamma}$ where $\gamma = 1.05\pm0.05$. The insets show direct relationship between $\tau^{(I)}$ and $\alpha$ (in strategy I) and $\tau^{(II)}$ and $p$ (in strategy II). }

\label{fig_2d_t}
\end{figure*}

\subsection{Three dimensional lattice}
Here the restaurants are assumed to be situated on the sites of a simple cubic lattice. $n_i(t-1)$ number of agents had chosen the $i$-th restaurant last evening (time $t-1$) then each of them chooses to go back to the same $i$-th restaurant with probability $[n_i(t-1)]^{-\alpha}$ next ($t$-th) evening and chooses one among the six of its neighbouring restaurants otherwise, for strategy I.  With strategy II, the probability for each of the $n_i(t-1)$ agents to go back to the $i$-th restaurant on the next ($t$-th) evening is $p$ for $n_i(t-1)>1$ and to any of the six neighbouring restaurants with probability $1-p$. After the system stabilizes i.e., when the utilization fraction $f(t)$ becomes independent of $t$, we note the steady state utilization fraction $f^{(I)}$ with strategy I or $f^{(II)}$ with strategy II and note the convergence time $\tau^{(I)}$ for strategy I and $\tau^{(II)}$ for strategy II, when $f(t)$ becomes $t$-independent. We find the power law fits for wastage fraction $(1-f^{(I)})\sim(1-f^{(II)})\sim(\alpha - \alpha_c)^{\beta}\sim(p_c - p)^{\beta}$ with $\beta = 0.87 \pm 0.05$ (see Fig.~\ref{fig_3d_f}) and $\tau^{(I)}\sim\tau^{(II)}\sim(\alpha-\alpha_c)^{-\gamma}\sim (p_c - p)^{-\gamma}$ with $\gamma =  1.11 \pm 0.07$ (see Fig.~\ref{fig_3d_t}) for both of the strategies I and II. All these studies are for $N\leq40^3$ and averaged over $4\times10^{5}$ runs.

\subsection{Two dimensional lattice}
Here the restaurants are assumed to be situated on the sites of a square lattice. $n_i(t-1)$ number of agents had chosen the $i$-th restaurant last evening (time $t-1$) then each of them chooses to go back to the same $i$-th restaurant with probability $[n_i(t-1)]^{-\alpha}$ next ($t$-th) evening and chooses one among the four of it's neighbouring restaurants otherwise, for strategy I. With strategy II, the probability for each of the $n_i(t-1)$ agents to go back to the $i$-th restaurant on the next $t$-th evening is $p$ for $n_i(t-1)>1$ and to any of the four neighbouring restaurants with probability $1-p$. After the system stabilizes i.e., when the utilization fraction $f(t)$ becomes independent of $t$, we note the steady state utilization fraction $f^{(I)}$ with strategy I or $f^{(II)}$ with strategy II and note the convergence time $\tau^{(I)}$ for strategy I and $\tau^{(II)}$ for strategy II, when $f(t)$ becomes $t$-independent. We find the power law fits for wastage fraction $(1-f^{(I)})\sim(1-f^{(II)})\sim(\alpha - \alpha_c)^{\beta}\sim(p_c - p)^{\beta}$ with $\beta = 1.0 \pm 0.05$ (see Fig.~\ref{fig_2d_f}) and $\tau^{(I)}\sim\tau^{(II)}\sim(\alpha-\alpha_c)^{-\gamma}\sim (p_c - p)^{-\gamma}$ with $\gamma =  1.05 \pm 0.07$ (see Fig.~\ref{fig_2d_t}) for both of the strategies I and II. All these studies are for $N\leq140^2$ and averaged over $16\times10^5$ runs.

\subsection{One dimensional lattice}
Here the restaurants are assumed to be situated on the sites of a linear chain. If $n_i(t-1)$ number of agents had chosen the $i$-th restaurant last evening (time $t-1$), then each of them chooses to go back to the same $i$-th restaurant with probability $[n_i(t-1)]^{-\alpha}$ next ($t$-th) evening and chooses one among the two of it’s neighbouring restaurants otherwise, for strategy I. With strategy II, the probability for each of the $n_i(t-1)$ agents to go back to the $i$-th restaurant on the next $t$-th evening is $p$ for $n_i(t-1) > 1$ and to any of the two neighbouring restaurants with probability $1-p$. It is straightforward to show the phase transition disappears and for any value of $\alpha (> 0)$ or $p < 1$ that $f$ is equal to unity and the convergence time $\tau$ is trivially dependent on $N$ (no critical slowing down or divergence near the critical point) as in other dimensions. This can be seen easily in the directed case, where each one chooses to hop to its right (or left) restaurants and the chain form a ring (with periodic boundary condition).

\section{Summary \& Discussion}

The KPR problem is an iterative, many choice, many player game, where the players try to learn from their past mistakes and from the publicly available information regarding the crowd-sizes of all the restaurants in the past, to choose one where she is expected to be alone today. We consider the cases here where the number ($N$) of agents (players) equals the number of resources (restaurants). It was shown \cite{chakrabarti2009kolkata} that no learning (random choices) leads to a societal resource utilization fraction $f$ = 1 - exp(-1) $\simeq$ $0.63$ (Eqs.~\ref{eq2.1},~\ref{eq2.2}). We study the collective learning induced phase transition in the KPR problem with (learned and mutually agreed) stochastic strategies I and II, where $f\to1$ at the respective critical points (with diverging convergence time due to critical slowing down). 
These are repetitive learning stochastic strategies, which are shown to reach eventually (after the convergence time) a
maximally utilized resource state which is best for most of the agents, who are not unique and every one in the game will
come in turn (stochastically) to this fortunate fraction (stochastic Nash type equilibrium). As mentioned already 
\cite{chakrabarti2009kolkata}, random choice by agents (myopic agents can be knocked out by others choosing that restaurant next
evening, as every player is equal), though still not one shot game, with utilization fraction can be analytically
estimated. Here we present improvements on those results.

We have studied here the the social utilization fraction $f(t)=1-\sum_{i=1}^{N} \frac{\delta(n_i(t))}{N}$, where $n_i(t)$ represents the number of customers who choose the $i$-th restaurant for $t$-th evening $(\delta(n) = 1$ for $n$ = 0 and $\delta(n) = 0$ for $n > 0)$ numerically (Eq.~\ref{eq1}). We considered here the case of equally ranked restaurants (and left the case where each of the restaurants have an unique rank and those ranks are agreed by all the agents for future study). The learning strategies employed by the agents here are strategies I and II respectively. Based on the observations from the previous studies~\cite{chakrabarti2017econophysics,ghosh2010statistics} that the probabilistic strategy to go back to the earlier chosen $i$-th restaurant on the next evening with probability inversely proportional to the crowd size $n_i(t)$ gives better success (compared to random choice), the strategies here (in both I and II) are developed such that the agents go back to the same restaurant on the next evening $(t+1)$ with probability $[n_i(t)]^{-\alpha}$ in I and with probability $p$ if $n_i(t) > 1$ in II, and chooses any of the $(N-1)$ other restaurants with rest of the probability (Eqs.~\ref{eq3.1},~\ref{eq3.2} and~\ref{eq4.1},~\ref{eq4.2}). As demonstrated, both the strategies (I and II) lead to identical statistics (singularities), though the strategy II is computationally little faster; see TABLE $\ref{tab:table1}$.

  \begin{table}[ht]
    \caption{ Steady state behaviours at $\alpha = 1.0$ and $p\to 0_+$ } 
    \label{tab:table1}
    \begin{tabular}{|c|c|c|c|c|} 
    \hline
     lattice dimension & \multicolumn{2}{c|} {\textbf{$\alpha = 1.0$}} & \multicolumn{2}{c|} {\textbf{$p = 0.05$}}   \\
    \cline{2-5}   & \textbf{$f$} & \textbf{approx. $\tau$} & \textbf{$f$} & \textbf{approx. $\tau$} \\
     
      \hline
	  $\infty d$ & $0.795 \pm 0.02$ & $5 \pm 1$  & $0.686 \pm 0.01$ & $2 \pm 1$ \\
	\hline     
	
      $3d$ & $0.814 \pm 0.02$ & $6 \pm 1$  & $0.628 \pm 0.01$ & $3 \pm 1$ \\
	\hline   
	
  $2d$ & $0.844 \pm 0.02$ & $12 \pm 1$  & $0.665 \pm 0.01$ & $4 \pm 1$ \\
	\hline   
	
    \end{tabular}
\end{table}

The steady state corresponds to the case where the average utilization $f(t)=f$ becomes $t$-independent and as shown here, $f = 1$ as $\alpha\to0_+$ in strategy I and $p\to1_-$ in strategy II, though the convergence time $\tau$ (to reach this $100\%$ utilization state) diverges there. Specifically we find: $f = 1-const\cdot(\alpha-\alpha_c)^\beta$ and $\tau\sim(\alpha-\alpha_c)^{-\gamma}$, $\alpha_c = 0_+$ for strategy I and $f = 1-const\cdot(p_c-p)^\beta$ and $\tau\sim(p_c-p)^{-\gamma}$, $p_c = 1_-$ for strategy II with $\beta = 0.80\pm0.05, 0.87\pm0.05, 1.0\pm0.05$ and $\gamma = 1.18\pm0.05, 1.11\pm0.05, 1.05\pm0.05$ in $d\to\infty$, $d\to3$ and $d\to2$ respectively for both of the strategies (I and II). It may be noted that the relaxation time $\tau$ diverges near critical point $(\alpha = \alpha_c$ or $p = p_c)$ which indicates full social utilization fraction $(f=1)$ is not achievable in the model at any practical convergence time limit. However in any phase transition, at the transition (critical) point, the relaxation (convergence) time $\tau$ diverges (often called critical slowing down). The observed singularity in the divergence here for $\tau$ confirm the phase transition behaviour in our KPR model near $\alpha_c$ or $p_c$ (with strategies I and II). To show the practical benefits of these strategies we also give here TABLE $\ref{tab:table1}$, showing the (very small) $\tau$ values for $\alpha(>>\alpha_c = 1)$ or $p(<<p_c = 0_+)$. Of course the social utilization fraction values obtained here are much less than maximum possible values of unity.

We hope, further studies (and applications) of this phase transition behaviour for the practical cases as suggested and considered in refs. \cite{park2017kolkata,yang2018mean,martin2017vehicle,martin2019} will contribute significantly in social science. 
It may be mentioned that in the dynamic matching of
car hire problem \cite{martin2017vehicle,martin2019}, extensive application of
strategy I (with $\alpha = 1$) suggested considerably
increased efficiency of optimization in the market,
while for avoiding crowd in job slot selection for
internet of things \cite{park2017kolkata}, strategy I was modified
to reward only the agents (here jobs) who come (or get 
thrown) alone for each restaurant (slot).
However such studies also face the formidable challenges to accommodate effectively the ranking of the restaurants, heterogeneity in their learning capacities to form queues spontaneously in finite convergence time.

\section{Data Availability Statement}
\noindent The data that support the findings of this study are available from the corresponding author upon reasonable request.

\textbf{\section*{Acknowledgments}}
\noindent We are thankful to Arnab Chatterjee for helpful suggestions. AS is grateful to Saha Institute of Nuclear Physics, Kolkata for hospitality and Asutosh College, Kolkata for support. BKC is grateful to J.C. Bose Fellowship (DST, Govt. of India) Research grant for support.


%

\end{document}